\documentclass[10pt,conference,letterpaper]{IEEEtran}

\usepackage[T1]{fontenc}
\usepackage[cmex10]{amsmath}
\usepackage{amssymb}
\usepackage{amsthm}
\usepackage{amsfonts}
\usepackage{algorithm}
\usepackage{algpseudocode}
\usepackage{algorithmicx}
\algrenewcommand\algorithmicindent{1em}%
\usepackage{cite}
\usepackage{textcomp}
\IEEEoverridecommandlockouts
\usepackage{times}
\usepackage[scriptsize]{subfigure}
\usepackage{paralist}
\usepackage{mathrsfs}
\usepackage{graphicx}
\usepackage{tabularx}
\usepackage{url}
\usepackage{mathbbol}
\usepackage{booktabs}
\usepackage{verbatim}
\usepackage{tikz}
\usepackage{balance}
\usetikzlibrary{arrows.meta,calc}

\usepackage{bm}            
\usepackage{xcolor}
\usepackage{multirow}
\usepackage{cuted}
\usepackage{stfloats}
\usepackage{changepage}

\begin{document}


\title{Evaluation and Optimization of Positional Accuracy for Maritime Positioning Systems}

\author{
     \IEEEauthorblockN{Atilla Alpay Nalcaci$^1$, Fidan Mehmeti$^1$, Wolfgang Kellerer$^1$, Florian Schiegg$^2$}
     \IEEEauthorblockA{$^1$ Chair of Communication Networks, Technical University of Munich, Germany\\
     $^2$ Covadonga GmbH, Germany
     \\\{atilla.nalcaci, fidan.mehmeti, wolfgang.kellerer\}@tum.de,  florian.schiegg@covadonga.eu}
}

\maketitle

\begin{abstract}
Navigation and trajectorial estimation of maritime vessels are contingent upon the context of positional accuracy. Even the smallest deviations in the estimation of a given vessel may result in detrimental consequences in terms of economic and ecologic quotients. To ensure an agile and precise environment for maritime vessel positional estimation, preexisting marine radar technologies can be utilized in a way that ensures a higher level of precision compared to GNSS-based identification and positioning.
In this paper, we present a positional  optimization for radar-based vessel navigation systems that utilize the installment of vessel detection sensors. The main objective of this research is to employ as fewer sensors as possible while preserving the attainable error threshold for positioning that is defined by International Maritime Organization (IMO). 
Our approach
leads most of the time to a positioning error of up to $5$\:m along shorelines and rivers and up to $50$\:m along open coastal regions. 

\end{abstract}


\section{Introduction}
\label{sec:introduction}
The importance and necessity of the maritime vessels are paramount in the context of global logistics and nautical transportation. According to the United Nations Conference on Trade and Development (UNCTAD), more than $80$\% of the international freights and shipments are transported using ships~\cite{un}. Currently, worldwide maritime traffic predominantly relies on Global Navigation Satellite Systems (GNSS). The positioning of the vessels is conducted in terms of the extraction and tracking of positional data in addition to the utilization of nautical charts and communication in a timely fashion~\cite{book1}.

GNSS systems that are employed by nautical vessels comprise of a transponder that is capable of transmitting the tracking and positioning information which allows practical facilitation by the responsible maritime authorities~\cite{shipnav1}. However, GNSS- and other wireless-based communication systems are immensely prone to numerous interference and malfunction scenarios. Main examples include spoofing and jamming attacks that may be conducted unintentionally or by third parties with malicious intent. Jamming is defined as the presence of a competing signal that prevents the decoding operation of a real satellite signal for the original GNSS receiver~\cite{gnss2}. Spoofing is the attempt to modify GNSS measurements with the intention of misinterpreting the signal as authentic by the original GNSS receiver~\cite{gnss5}. To that end, the vulnerability of GNSS-based technologies—the most prominent being the Global Positioning System (GPS)—is referred to in numerous studies~\cite{gnss2,gnss3,gnss4} in terms of the proliferation and suppression of GPS jamming attacks from the maritime perspective.

To that end, the implementation of a back-up system is crucial for enhancing the nautical safety in the event of the existing systems not being operational or compromised for an uncertain amount of time. A radar-based navigation system, however, does not pertain to the issues mentioned above. In particular, a radar-based positioning system aims to estimate the range of a vessel~\cite{radarbook}. The system consists of a radar antenna (an antenna for each given vessel) and sensors along a given region, typically a fjord or a shoreline. A directional radar antenna transmits the signal to the desired location~\cite{radar1}. Once the radar beam reaches a sensor, the timestamp value of when a sensor is hit with a radar beam is logged, which is then utilized in various positioning techniques, ultimately yielding the approximate position of the vessel.

\textcolor{black}{
The radar-based vessel identification, tracking, and guiding system presented in this paper employs multilateration in order to locate the vessels in a given region. Multilateration is a technique used to determine the location of an object by measuring the time it takes for signals to travel from the object to multiple known locations~\cite{lclz1}. Only the sensor positions and timestamp values of the radar beams hitting those sensors are known to our system. An infrastructure capable of successfully locating maritime vessels as a backup solution is shown and used relying on our approach.}

\section{Background and Related Work}
\label{sec:related_work}


Aftanas \emph{et al.}~\cite{b1} propose a method for analyzing target positioning accuracy for $M$-sequence UWB radar systems under ideal conditions. It has been shown that the target position estimation error is caused by quantization effects at time-of-arrival (TOA) measurement by using $M$-sequences. Findings suggest that the largest target position estimation errors are located along the straight lines between TX and all RX antennas, but only behind antennas excluding the area between them. The best system accuracy is obtained when the distance between antennas is equal to $5$\:m. More RX antennas improve the accuracy of the UWB radar system. 
Results suggest the potential use of this approach under more realistic scenarios. 
It can also be applied in sensor network design for detection, positioning and tracking of selected targets.

In their work, Bishop \emph{et al.}~\cite{b5} aim to identify relative sensor-target geometries which result in minimizing the uncertainty ellipse. Optimal sensor-target geometries for range-only, time-of-arrival-based and bearing-only localization are identified and studied, and a direct and rigorous characterization of the relative sensor-target geometry were given. The characterizations are given in terms of the potential localization performance of unbiased and efficient estimators. The contributions provide an explicit and measurable connection between the sensor-target geometry and the chosen measure of localization performance, i.e., the area of the uncertainty ellipse. A number of necessary and sufficient conditions on the sensor-target angular positions were given, 
\textcolor{black}{which then minimizes the unbiased and efficient target uncertainty ellipse area that provides the aggregate of the points which lie within the defined ellipses bounds}. It is also shown that an optimal sensor-target configuration is not unique in a general sense.

\textcolor{black}{Finally, according to Naus \emph{et al.}~\cite{b3}, a premise for adopting autonomous methods in terms of navigation systems is paramount for providing alternatives to Global Navigation Satellite System (GNSS). The accuracy analysis carried out  with respect to the simulated radar positioning in relation, marks the validity of the research.} The positional accuracy on fairways was better than $6.5$\:m. Then, the association of extracted characteristic radar echoes with navigation marks found in an ENC was made. Ultimately, radar imaging is clearly a source of information on the basis of which it is possible to determine the position of the ship. The methods most commonly used for this purpose are those that individually use bearing and distance measured to the navigation marks with known coordinates visible on the radar image. The article presents the method of position determination based on information on the position of navigation marks obtained directly from an ENC, thus eliminating the effect of map projection distortions completely.

\section{Methodology}
\label{sec:method}
\subsection{Circumambient Circle Generation}
The system utilizes the locations of maritime radar detectors and timestamp values that are obtained when the receivers are hit by an incoming radar beam. Calculations also include a radar period value that defines the full rotation of a radar antenna in the given time unit. Let $\alpha$ be the angle at the initial measurement of vessel location to the sensors $s_i$ and $s_j$. For two sensors that are hit at different times, and the radar period value $P$, the angle to the both sensors from the initial measurement of the vessel position can be calculated.


Note that sensor labeling occurs based on the timestamp value that is recorded when the corresponding sensor is hit. The system is capable of labeling the sensors on account of when the first timestamp log occurs. In addition, for the rest of this paper, the timestamp difference of $t_i$ and $t_j$ for given sensors $s_i$ and $s_j$ will be depicted by the variable $\tau_{ij}$. The formula then reduces to
\begin{equation}
    \alpha(\tau_{ij}, P) = \frac{2\pi\tau_{ij}}{P}. \label{eq:alpha_angle}
\end{equation}

A preliminary distance calculation to the vessel is paramount since without it, the vessel may be lying at any point with the measured angle value. To that end, a circle construction is carried out where the circumference of the generated circle intersects with the sensor positions. Let $s_i = (x_i, y_i)$ and $s_j = (x_j, y_j)$ be the positions of the two sensors. Accordingly, $d_{ij}$ is the distance between these coordinates.


The radius of the circumambient circle to the sensors with the angle $\alpha$ can then be calculated as
\vspace{-9pt}

\begin{equation}
    r(\alpha, d_{ij})=\frac{d_{ij}}{\sqrt{2(1 - \cos 2 \alpha)}}. \label{eq:radius}
\end{equation}
Subsequently, the center point of the circle circumambient to the sensor positions and the given radius function can then be obtained from
\begin{equation}
    P_{c_{ij}} = \Bigg(\begin{array}{cc} x_i \\ y_i \\ \end{array}\Bigg) + \Bigg(\frac{r}{d_{ij}} \cdot \Bigg(\begin{array}{cc} \sin\alpha & -\cos\alpha\\ \cos\alpha & \sin\alpha \\ \end{array}\Bigg)\Bigg) \Bigg(\begin{array}{cc} x_j - x_i \\ y_j - y_i \\ \end{array}\Bigg). \label{eq:circle_center}
\end{equation}

Now, the problem statement is nondeterministic as every point on the circumference of the circle is included in the solution set. Accordingly, the system requires a definition of at least another spanning circle, in terms of another sensor. Let us introduce another sensor to the system. Symmetrically, the sensor that is defined will now be  $s_j = (x_j, y_j)$ whereas the sensor from previous calculations under the label  $s_j$ will now shift to $s_i = (x_i, y_i)$, to delineate the receiver setup. Following that, the system generates two additional spanning circles whose circumference pass through a joint intersection that is used as an initial estimation of the vessel origin.
\begin{figure*}
        \centering
        \includegraphics[width=7in]{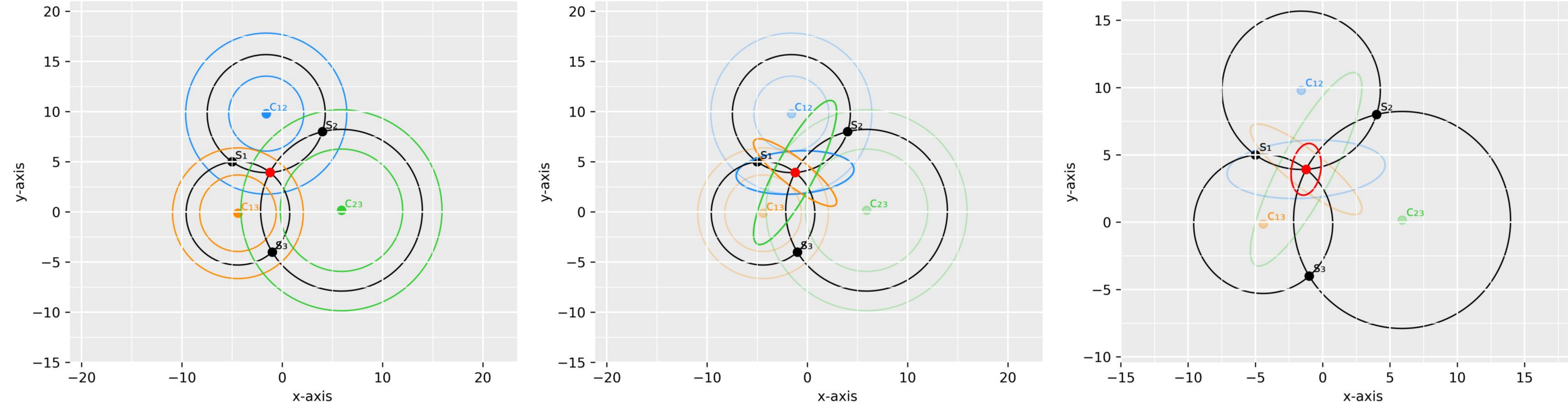}
        \vspace{-12pt}
        \caption{(left) Upper- and lower-bounding circumambient circles, (middle) uncertainty area generation pertaining to distinct circumambient circle, (right) Final confidence region covering approximately 95\% of all potential vessel positions.} 
        \label{fig:three_combined}
        \vspace{-12pt}
\end{figure*}

\textcolor{black}{In each time step, the circle generation scheme from (\ref{eq:circle_center}) is carried out where the corresponding radius, distance, and angle entries are measured. In order to conduct an uncertainty propagation in terms of radius, timestamp values gathered from the sensors are cross-referenced with respective ground truth values.} With $\tau_{ij}$ depicting the measured time difference between two sensors, let $t_i'$ and $t_j'$ denote the respective ground truth timestamp entries for sensors $s_i$ and $s_j$. \textcolor{black}{Denoting the time difference as $\Delta t_{ij}$, the time difference between the measured and ground truth timestamp values then becomes}
\begin{equation}
    \Delta\tau_{ij} = \Delta\Delta t_{ij} = (t_j - t_i) - (t'_j - t'_i).
\end{equation}

The incorporation of the term $\Delta\tau_{ij}$ is to describe the measurement error in terms of timestamp measurements. Subsequently to this deviation, a refinement pertaining to the radius estimation of the system is carried out. By substituting the computed angle $\alpha$ within the radius calculation (\ref{eq:radius}) 
with associated parameters $\tau_{ij}$ and $P$, the radius estimation is then 
\begin{equation}
    r(\tau_{ij}, d_{ij}, P)=\frac{d_{ij}}{\sqrt{2 - 2 \cos {(\frac{8\tau_{ij}\pi}{P})}}}. \label{eq:radius}
\end{equation}

The radius error pertaining to each circumambient circle can be computed as a derivation of the radius with respect to timestamp difference $\tau$ and a combination of this derivation with timestamp measurement error $\Delta\tau_{ij}$. Let $\Delta r$ represent the difference in radius that will be extracted upon the association of ground truth values:
\vspace{-9pt}

\begin{equation}
    \Delta r=r({\displaystyle \tau_{ij} +\Delta } \tau_{ij} ) -r({\displaystyle \tau_{ij} }). \label{eq:delta_r}
\end{equation}

This yields the uncertainty propagation in terms of the distinct circumambient circles. In order to compute the incorporation of this difference to each radius estimation, an approximation of $\Delta r$ must be carried out in terms of the real time difference $\tau_{ij}$,
\vspace{-9pt}

\begin{equation}
    r({\displaystyle \tau_{ij} +\Delta } \tau_{ij} ) -r({\displaystyle \tau_{ij} }) \approx ||\frac{dr}{d\tau_{ij} }||{\displaystyle \Delta } \tau_{ij}. \label{eq:derive}
\end{equation}

To estimate this difference, the computation of the first derivative of the radius function with respect to $\tau$ is performed. In particular, by substituting (\ref{eq:derive}) with the extended radius formula (\ref{eq:radius}), the deviation in circle radius is then written as
\begin{equation}
    \Delta r\approx \frac{-d_{ij}}{P}\cdot{\displaystyle \frac{\left({\displaystyle 2\pi}\sin\left(\frac{4{\displaystyle \pi } \tau_{ij} }{P} \right)\right)}{\sqrt{\left( 2-2\cos\left(\frac{4{\displaystyle \pi } \tau_{ij}}{P}\right)\right)^{3}}} \cdot \Delta } \tau_{ij}, \label{eq:delta_r_extended}
\end{equation}
yielding the deviation from the true calculation of the spanning circle. Since a deviation may occur in both directions, a diminutive confidence region is formed towards the measurement of the vessel positional solution set. The incorporation of the delta value to the existing circumambient circle setup reveals the approximate confidence region, which is detailed in the next subsections. Fig.~\ref{fig:three_combined}(left) illustrates the delta value corresponding to theradius estimation applied to the anterior circumambient circle illustration.

\subsection{Confidence Ellipse Approximation}
The main goal of the circumambient circles is to transpose the positioning estimation from a multiangulation setup to a multilateration. Inherently, circles depict virtual anchors. \textcolor{black}{As the system calculates the corresponding confidence regions of sensors in tuples, the final region represents the confidence ellipse which encapsulates approximately $95$\% of all potential vessel points within a given area.} In order to estimate the grand area for every sensor tuple, the system calculates the semi-minor, semi-major, and orientation of the ellipse to be formed. Let $a_{ij}$ denote the semi-minor axis of the ellipse formed out of sensors labeled $i$ and $j$. Let $b_{ij}$ denote the semi-major and $\phi_{ij}$ denote the orientation of the ellipse. The center of the ellipse is the initial intersection of the two circular areas formed out of the respective sensors. The values of semi-minor and semi-major axes are given by 
\vspace{-9pt}

\begin{equation}
a_{ij} = \Delta r_{ij} \label{eq:semi_n}, \quad b_{ij} =r_{ij}, 
\end{equation}
where $\Delta r_{ij}$ is the measured deviation from ground truth values, and $r_{ij}$ is the radius of the circumambient circle corresponding to sensors $i$ and $j$. Calculating the reverse tangent of the slope of semi-major axis, the system is capable of computing the orientation of the ellipse. Let $(x_{b_{ij}}, y_{b_{ij}})$ and $(x'_{b_{ij}}, y'_{b_{ij}})$ be two points through which the semi-major axis $b_{ij}$ pass. One of the points, denoted as $(x_{b_{ij}}, y_{b_{ij}})$, is the intersection point of the generated circumambient circles. The other point, represented by $(x'_{b_{ij}}, y'_{b_{ij}})$, is the center of the corresponding circle $c_{ij}$ to the ellipse that will be generated. The system utilizes the point-slope form of a line to calculate the slope and further incorporate into the ellipse orientation as
\begin{equation}
{\phi _{ij} = \arctan(\frac{y_{b_{ij}} -y'_{b_{ij}}}{x_{b_{ij}} -x'_{b_{ij}}})},
\label{eq:ellipse_angle}
\end{equation}
which effectively yields the initial confidence regions pertaining to distinct circular vessel frames. Fig.~\ref{fig:three_combined}(middle) represents an exemplary scenario that exhibits three ellipses formed out of a 3-sensor setup. The ellipses that are formed describe the dispersion of data points around the mean. The outlined area provides the quantification of how much individual data points have deviated from the mean. The covariance matrix corresponding to each ellipse, represented as $\Upsilon_{ij}$ for sensors $s_i$ and $s_j$ provides a summary of variances for each estimated confidence region. The diagonal elements of a given covariance matrix represent the variance of individual variables. The off-diagonal elements represent the variances between pairs of variables. For a circle $c_{ij}$ with center point $P_{c_{ij}}$, the spread of data points around the mean for $x$ and $y$ axes becomes
\vspace{-9pt}

\begin{equation}
    {\displaystyle \sigma _{x_{i} x_{j}} =b_{ij}^{2}\cdot \cos( \phi _{ij})^{2} +} a_{ij}^{2}\cdot \sin({\displaystyle \phi }_{ij})^{2},
\end{equation}
\begin{equation}
    {\displaystyle \sigma _{y_{i} y_{j}} =b_{ij}^{2} \cdot\sin( \phi _{ij})^{2} +} a_{ij}^{2} \cdot{\displaystyle \cos( \phi _{ij})}^{2},
\end{equation}
\begin{equation}
    {\displaystyle \sigma }_{x_{i} y_{j}} = {\displaystyle \sigma }_{x_{j} y_{i}} = \left( b_{ij}^{2} -a_{ij}^{2}\right)\cdot \sin({\displaystyle \phi }_{ij})\cdot {\displaystyle \cos( \phi _{ij})}.
\end{equation}
Following that, from the primary corollary regarding the quantification of the relationships between multivariate variables, the open form of the appurtenant covariance matrix $\Upsilon_{ij}$ is
\begin{equation}
    \Upsilon_{ij} = \Bigg(\begin{array}{cc} \sigma_{x_{i} x_{j}} & \sigma_{x_{i} y_{j}}\\ \sigma_{y_{i} x_{j}} & \sigma_{y_{i} y_{j}} \\ \end{array}\Bigg).
\end{equation}

The system incorporates the impeding covariance values to the final confidence region by means of combining the means of distributions within the configuration in order to represent the non-normalized intersection of the estimated populations of vessel positions. Let $\Upsilon'$ depict the covariance matrix of the final confidence ellipse for a given setup. The derivation of the matrix entries of $\Upsilon'$ is done on the basis of distinct covariance matrix values gathered from the initial ellipse formations. For a system comprising three sensors, $s_i$, $s_j$, and $s_k$, let $\Upsilon_{ij}$ and $\Upsilon_{jk}$ correspond to covariance matrices. For configurations encompassing three sensors or more, the incorporation of all covariance matrices is performed as
\vspace{-9pt}

\begin{equation}
    \Upsilon ^{'} =\Upsilon _{ij} -{\displaystyle \left( \Upsilon _{ij} \cdot ( \Upsilon _{ij} +\Upsilon _{jk})^{-1}\right) \cdot } \Upsilon _{ij}, \label{eq:conf_reg_final}
\end{equation}
where the final matrix describes the confidence ellipse. At each time step, the most recent covariance matrix is combined with the estimated combined covariance matrices (represented by $\Upsilon _{ij}$), and the inverse of the product is then multiplied by the most recent covariance matrix again to extract the multivariate distribution regarding the transformations amongst distinct sensors. Fig.~\ref{fig:three_combined}(right) shows the completed confidence region generation scheme of the 3-sensor setup provided so far. The generation scheme of the confidence region is carried out through the utilization of eigenvalues that represent variances along the corresponding eigenvectors. Eigenvectors represent the directions of maximum variances. The final region represents an encapsulation of the solution set that contains approximately $95$\% of the estimated vessel positions, i.e., $95$\% of all points at which the vessel can potentially be located. The outcome is the trajectorial examination of vessels for each time step. The latter is defined depending on the arrival of radar beam to the maritime radar detectors. The system performs the computation of distinct sensor tuples to the received timestamp values. A given position is further improved once other sensors receive a radar beam commensurate to the same vessel identification. 

\subsection{Model Verification}
\begin{figure}
    \centering
    \includegraphics[width=3in]{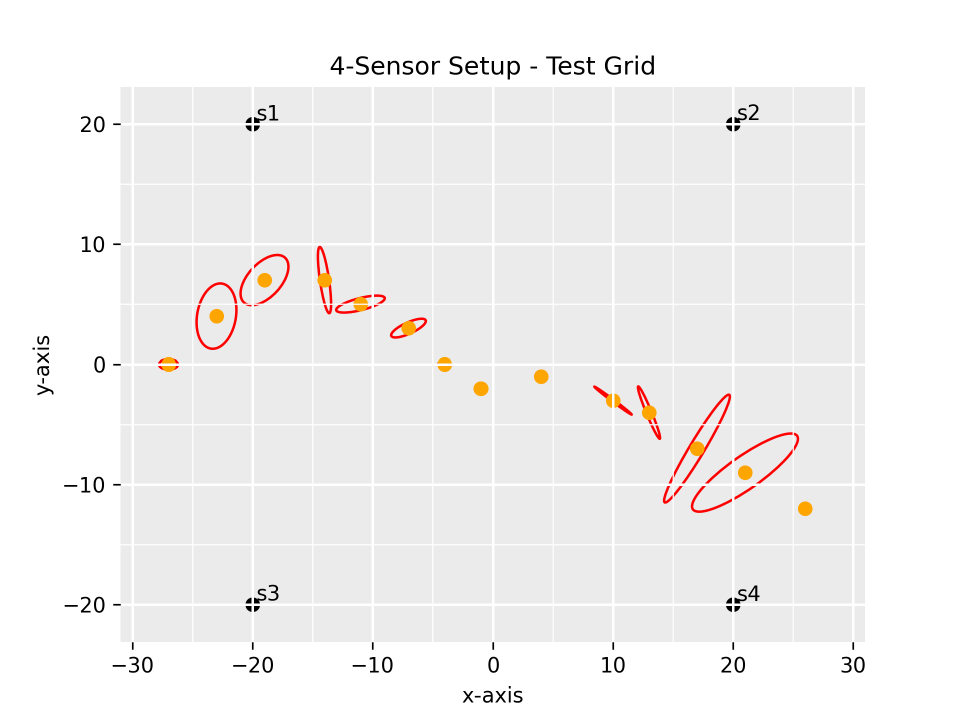}
    \vspace{-12pt}
    \caption{An example 4-sensor setup consisting of a singular vessel. Orange dots represent vessel route over time. The confidence region for each time step is illustrated with red ellipses.}
    \label{fig:figure_ellipses_over_time}
    \vspace{-12pt}
\end{figure}

The verification 
of the mathematical model was conducted in two phases. The initial phase provides the generation of a test environment where sensor positions, timestamps with corresponding ground truth values, and vessel grid were defined manually. For a given list of sensor positions, our approach calculates the bounding circles for each combination of sensor tuples. Then, upper and lower bounding circles are generated with their following confidence ellipses. The final confidence region is then derived with reference to the provided methodology. The full deployment of the system internally concentrates on the final confidence ellipse in order to elucidate the change in positioning error over time. Fig.~\ref{fig:figure_ellipses_over_time} depicts a sample setup carried out in the test environment with a predefined vessel and four sensors.

In each assessment, the confidence region is calculated for the given time step. In order to incorporate further analysis and topological evaluation, the implementation of the initial test environment is rudimentary with regard to exploring system specifications. By virtue of the initial tests, it is shown that the system is capable of functioning on a fundamental level for the parameter values that are required for its practical purposes.

The second phase consists of a bottom-up approach to the methodology for the substantiation of the positioning algorithm. A reverse approach in terms of validating mathematical precision throughout a given environment was carried out. The goal is to feed to the algorithm each grid point as the vessel position with the aim of acquiring a heat map that represents the prediction accuracy regarding positioning of each grid entry. This allows a thorough analysis of a given region, in addition to validating the stability of the provided methodology behind the vessel positioning. The system estimates the positioning error by reverse calculating the angle of the vessel to every sensor tuple. \textcolor{black}{For the verification, the vessel position is known internally. Let $v$ denote the vessel position for a given time step instance.} By utilizing the cosine law using (\ref{eq:alpha_angle}) we can calculate the inverse angle value as
\begin{equation}
    \alpha =\arccos\left({\displaystyle \frac{||s_{i} -s_{j} ||^{2} -||s_{i} -v||^{2} -||v-s_{j} ||^{2}}{-2 \cdot||s_{i} -v|| \cdot ||v-s_{j} ||}}\right), \label{eq:arccos_eq}
\end{equation}
This is the inverse angle approximation per vessel position. Once the system establishes a value for the angle using (\ref{eq:arccos_eq}), the boundary circle and the confidence ellipse generation schemes are carried out. The cross-evaluation of the system enables the posterior analysis of the test grid with real-world statistics. For a given region, our system is further enhanced with a focus on adapting vessel and sea map scales to conventional UTM and distance units.

\section{Optimization Setup}
\label{sec:optimization}


\subsection{Topological Analysis}
The major predicament in the context of regional topology is to scrutinize the inverse cosine of a given position to sensors. In cases when the vessel aligned with sensors, the corresponding circle radius value (\ref{eq:radius}) tends to infinity. Accordingly, the confidence ellipse computations yield predictions of unmeasurable positional error values.
\begin{figure}
    \centering
    \includegraphics[width=2.9in]{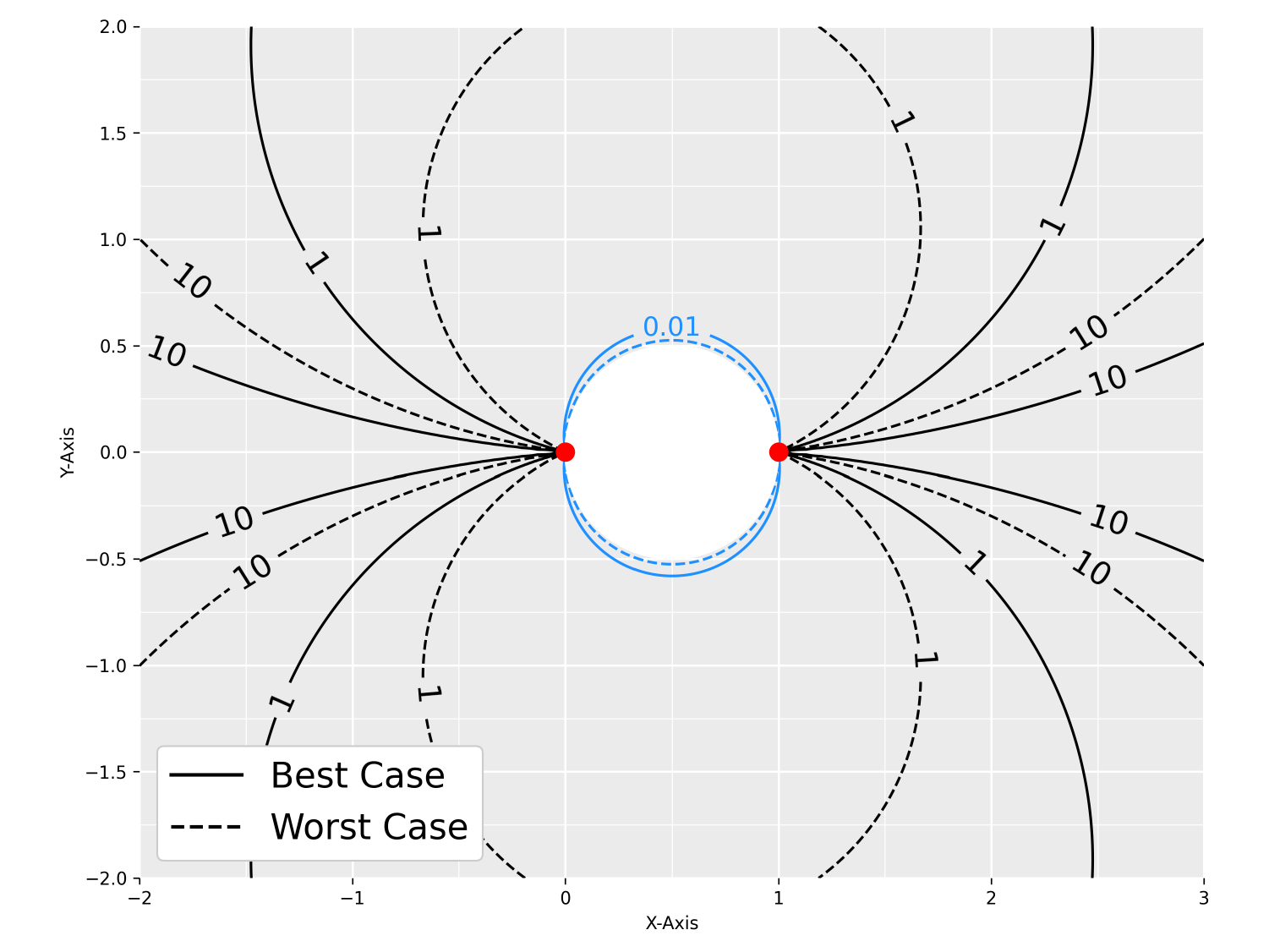}
    \vspace{-9pt}
    \caption{The error in the radius based on two receivers and the distance to the ship. The two MRD receivers 
    are marked as red dots. Solid contour lines depict the best case scenario to corresponding radius error propagation; dashed lines depict the worst case.}
    \label{fig:error_radius_worstbest}
    \vspace{-12pt}
\end{figure}

\begin{figure}
    \centering
    \includegraphics[width=2.9in]{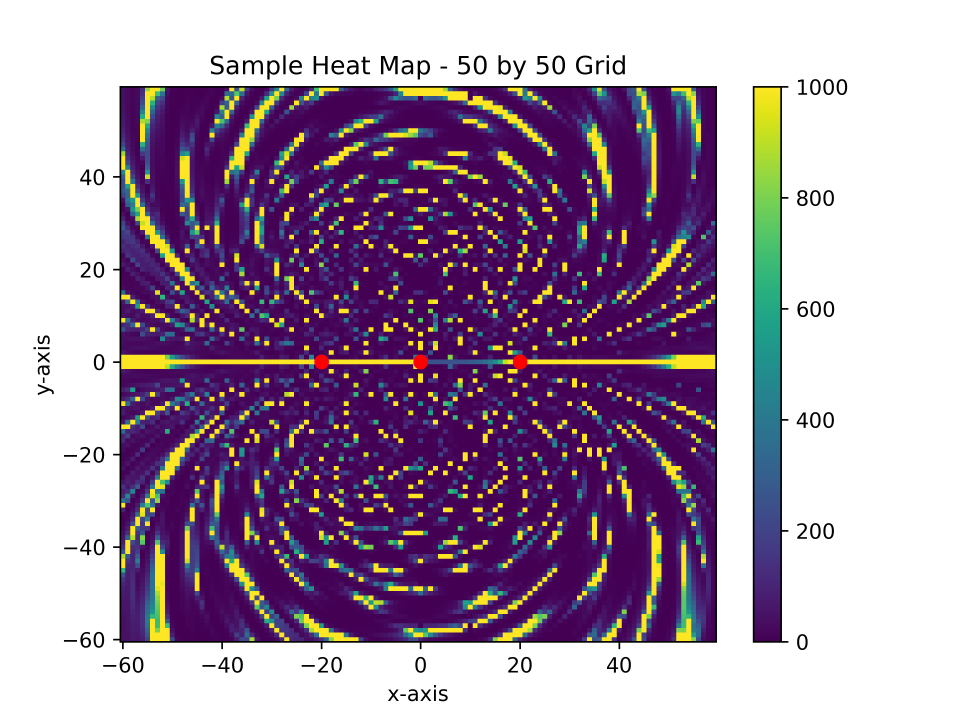}
    \vspace{-12pt}
    \caption{Three sensors placed with equidistant horizontal alignment. Sections between the sensors elongate on the same horizontal axis yield higher accuracy error compared to sections with elevated angle to the sensors.}
   \label{fig:figure_testenv_horizontalerror}
    \vspace{-12pt}
\end{figure}

For an augmented examination of positioning adeptness, Fig.~\ref{fig:error_radius_worstbest} illustrates the contour plot representation of a sample setup consisting of two sensors. Contour plots represent the radius error in terms of distance between the vessel and sensors. Contours that correspond to very low positioning error values are delineated with solid lines, and contours with high positioning errors are depicted in dashed lines. The areas surrounded by the contour plots represent the radius error ratio internal to that area. The goal is to achieve sensor localization scenarios that enable enhanced coverage for contour regions with smaller radius error propagation. The uncertainty propagation given by (\ref{eq:radius}) and (\ref{eq:delta_r}) are further observable in addition to contour plot representation of separate regions. In particular, the vessels that are on, or very close to the uniform alignment of two sensors, yield a difference in the radius ($\Delta r$ in (\ref{eq:delta_r_extended})). As defined in (\ref{eq:semi_n}), appertaining values in uncertainty ellipses with very high coverage lead to high positional error measurements. An example implemented in the test environment is shown in Fig.~\ref{fig:figure_testenv_horizontalerror}. The positional error yields values above the range $[0, 1000]$ in the areas located between the sensors. 
As (\ref{eq:delta_r_extended}) diverges to $\infty$ as the cosine term in the denominator approaches $1$, certain localization schemes should be avoided in order to minimize areas with unmeasurable positioning errors. Fig.~\ref{fig:figure_testenv_horizontalerror} represents the blatant example when all sensors are aligned to the same horizontal axis, where regions in-between sensors have exponentially increasing positioning error values. The placement and alignment geometries should satisfy the prospective vessel positioning irrespective of diagonality.

\begin{table}[t]
  \centering
  \caption{Energy consumption for one maritime radar detector (courtesy of \textit{Covadonga GmbH})}
  \vspace{-9pt}
  \begin{tabular}{|c|c|c|c|}
    \hline
     \textbf{Energy Consumption}&\textbf{Daily}& \textbf{Weekly} &\textbf{Monthly}\\
    \hline
     \textbf{Watt-hour} & 216 Wh & 1512 Wh & 6480 Wh\\ \hline 
     \textbf{Kilowatt-hour} & 0.216 kWh & 1.512 kWh & 6.48 kWh\\\hline
 \textbf{Joules} & 32400 J & 226800 J & 972000 J \\\hline
  \end{tabular}
\label{tab:power_consumption_rate}
\vspace{-12pt}
\end{table}

\subsection{Expenditure Evaluation}
The total expenditure is examined in terms of transmission of overall data and energy consumption over time. Depending on a vessel positional entry for a given time, not all sensors can be incorporated equally. Especially among areas with very high or very low positioning error, the incorporation of an additional sensor may be redundant. Before conducting comparative analyses between sensors for distinct positional scenarios, a breakdown of energy expenses and data quota tariffs are shown. Table~\ref{tab:data_transmission_rate} represents the data transmission ratio of one MRD sensor, subdivided into distinct cumulative time intervals. The average values are extracted with respect to estimations of real-world  measurement campaigns. Worst case scenarios presume consistent data flows of $4$\:Mbps over time (without disconnection). Measurements may vary depending on the amount of vessels and ascend of the region.

\begin{table}[t]
  \centering
  \caption{Price of data tariffs defined by \textit{Telefónica Telecommunications Company AG}}
  \vspace{-9pt}
  \begin{tabular}{|c|p{1.1cm}|p{1.7cm}|p{1.5cm}|c|}
    \hline
    \textbf{Data Tariff} & \textbf{Package Price} & \textbf{incl. Dynamic Data Pooling} & \textbf{incl. Additional connectivity} & \textbf{Total} \\
    \hline
    Per 100MB & 4.00 € & + 0.80 € & + 1.80 € & 6.60 € \\
    \hline 
    Per 1GB & 7.00 € & + 1.40 € & + 3.50 € & 11.90 € \\
    \hline
    Per 5GB & 17.50 € & + 3.30 € & + 8.75 € & 29.55 € \\
    \hline
  \end{tabular}
\vspace{-12pt}
  \label{tab:telefonica_price_range}
\end{table}

\begin{table}[t]
  \caption{Data transmission rates pertaining to a single MRD device (courtesy of \textit{Covadonga GmbH})}
  \vspace{-9pt}
  \centering
  \begin{tabular}{|c|c|c|c|}
    \hline
     \textbf{Time Interval}&\textbf{Minimum}& \textbf{Average}& \textbf{Worst Case}\\
    \hline
     \textbf{Per Day}&15.64 MB& 62.573 MB& 432 GB\\ \hline 
     \textbf{Per Week}&109.48 MB& 438.011 MB& 3024 GB\\
    \hline
 \textbf{Per Month}& 469.2 MB& 1.87719 GB& 13.16032 TB\\\hline
  \end{tabular}
\vspace{-12pt}
\label{tab:data_transmission_rate}
\end{table}

The quota tariffs that are used by Covadonga MRD devices are defined by \textit{Telefónica Telecommunications Company AG}. Prices correspond to standard IoT directory with corresponding expense rates. For each MRD receiver, transmission rates akin to overall expense are derived in terms of the provided data arrangement accordingly. Table~\ref{tab:telefonica_price_range} depicts the price range defined for a standard IoT device and its associated peripherals, with included dynamic data pooling and connectivity prices. Expense portfolios related to Germany are taken into account.

Consumption ratio is examined containing the totality of infrastructure, in particular, central processing unit and the sensors that incorporate to the system as a whole. Table~\ref{tab:power_consumption_rate} represents the power consumption with reference to the system. Conversions are made based on a single MRD consuming $9$\:W. The consumption ratio is depicted in addition to further denotation of the generic power consumption for progressive time intervals. Finally, the cost of installation is added to the expense, and is evaluated with respect to the energy and data rates provided so far. This value is inherently equal to the number sensors multiplied by the installation cost for one MRD receiver. On the Covadonga side, installment costs between $2,000$-$10,000$\:€, depending on the exact peripheral specifications of hardware and region to be installed. Installment valuation includes the expense pertaining to hardware peripherals that are embedded within a receiver.

The definition of the optimization setup is carried out in the context of decreasing the total expenditure of a proposed sensor infrastructure while keeping the positioning error under a given threshold. In terms of actual definition of the threshold - IMO features stringent requirements with respect to per-vessel positioning accuracy. For coastal waterways, defined as the interconnected bodies of water located near coastlines, the positioning error is expected to be less than $10$\:m~\cite{imolimit}.
In ports and harbors, the positioning error is postulated to be at most $1$\:m, and $0.1$\:m for docked vessels~\cite{imolimit}. Limitations pertaining to specific measurement error thresholds widely vary in a given region intrinsically in addition to the variation of the regions grand location. Once positioning error of a given timestamp is measured, the system is optimized in terms of detecting the sensor(s) with the least sensors to the measurement.

\subsection{Optimization Setup}
\textcolor{black}{The estimation cost of an infrastructure is examined under installation and operation of the given sensor network.} The expenditure model presented in this section does not incorporate potential system malfunctions or maintenance costs over time because these values are highly susceptible to alterations depending on the region of origin.

For a given region, once the overall number of sensors that will sustain optimal coverage is determined, the total installation cost is then d
    $I_{\text{total}} = N \cdot L,$
where $N$ denotes the number of sensors and $L$ denotes the installation cost per sensor. Since the same sensor model is utilized in a given region, the corresponding cost value is set for each distinct sensor installation. On the Covadonga side, the price of an MRD is rated between $2,000$\:€ to $10,000$\:€, depending on the equipment specifications. An MRD includes base radar receiver capabilities, an embedded interface, and a CPU adept for radar pulse detection and characterization. The operational cost of the infrastructure should also be incorporated to the expenditure analysis. The conservation of a sensor-based vessel identification infrastructure involves two operational perspectives, \emph{energy} and \emph{data}. Let $e_i(t)$ define the energy cost per sensor over time, 
whereas $\omega_1$ represents the corresponding energy unit conversion. Let $tr_i(t)$ define the transmission data cost per sensor over time, 
where $\omega_2$ represents the corresponding transmission rate unit conversion. The efficacy of the overall infrastructure is also crucial in order to minimize the use of resources. Accordingly, the system is able to detect and integrate if a sensor has not been operational (e.g., it has not received any signal). In such an instance, a sensor is able to switch to standby mode, where power consumption becomes negligible. In order to incorporate this feature, we define a binary decision variable $op_i(t)$ depicting whether the subject sensor was operational during the time frame $t$. We have
\vspace{-9pt}

\begin{equation}
    O_{total}( t) =\sum\limits _{n=1}^{N}[{\displaystyle op_{n}( t) \cdot ( e_{n}( t) \cdot \omega _{1} +tr_{n}( t) \cdot \omega _{2})]},
\end{equation}
where $O_\text{total}$ denotes the operation cost. Both models are then combined, generating the final expenditure of a given infrastructure. Since the calculations are carried out for a given time frame $t$, the final analytical model corresponding to total expense must also contain the overall time horizon, typically calculated per week or per month. In order to define the system over the time horizon, expenses that have been provided so far are defined over an integral. 
This can be approximated by
\[
\int_{0}^{T} O_{total} \, dt \approx \sum\limits _{t=0}^{T} O_{total}(t).
\]

Finally, by replacing the integral term with the discretized form and adding the fixed installation cost for the system, the final model pertaining to the expenditure evaluation becomes
\vspace{-9pt}

\begin{equation}
    Exp_{total} =\left[\sum\limits _{t=0}^{T} O_{total}(t)\right] +I_{total}. \label{total_exp}
\end{equation}
$Exp_{total}$ denotes the total expenditure of the infrastructure. While time frame calculations are made for varying target-sensor transmissions and instances, the main goal regarding the cost model is to minimize the overall expenditure with respect to operation and installation costs. Correspondingly to the confidence ellipse equations presented in Section~\ref{sec:method}, the mathematical setup of the optimization scheme is deducted. For a final covariance matrix $\Upsilon$ representing the multivariate distribution of vessel locations for a given timestep is 
\vspace{-9pt}

\begin{equation}
    \Upsilon = \Bigg(\begin{array}{cc} \sigma_{xx} & \sigma_{xy}\\ \sigma_{yx} & \sigma_{yy} \\ \end{array}\Bigg).
\end{equation}

The positioning error in floating point values is computed using the diagonal entries. Let $err_{s_{1\ldots N}}$ define the positioning error value with respect to sensors $1,...,N$. 
The positioning error in each time step is then 
$$
err_{s_{1\ldots N}} =\sqrt{{\displaystyle  \sigma_{xx}^{2} + \sigma_{yy}^{2}}}.
$$

Subsequently, the goal is to minimize the overall expenditure (\ref{total_exp}) while keeping the $err_{s_{1\ldots N}}$ under the threshold $\Omega$:
\begin{equation}
\begin{aligned}
\min \quad & Exp_{total} \\
\text{s.t.} \quad & err_{s_{1\ldots N}} \leq \Omega.
\end{aligned}
\end{equation}

The optimization of sensor clusters is carried out in connection with cross referencing the measured positioning error throughout the installed sensor within a given region. The main idea is to first ascertain that the positioning error is under the desired threshold value and then we utilize the sensor combination that is capable of achieving a valid positional error, ultimately employing fewer sensors. The resulting values are dedicated to consolidate tidal regions with respective sensor functionality schemes.
\section{Performance Evaluation}
\label{sec:testsresults}


\subsection{Rhine Measurement Campaign}
The first measurement campaign took place eastern of Germersheim, Germany. The covered region spans approximately $500$\:m in width and $1.5$\:km in height. Within the grid, six distinct MRD sensors were placed, conducting peak detection and ranging for three vessels. Each positional entry is labeled depending on the order of that entry in time to prevent clustering within the time series representations. The grid construction was carried out by detecting the sensor with farthermost latitudinal entries for the delineation of north and south boundaries, and longitudinal entries for the delineation of east and west boundaries of the grid. A positional margin value of $250$ grid cells was added to maintain a perimeter around the sensor. Subsequently, the defined grid was partitioned into $800$ even cells. For this scenario, the width of one grid cell is approximately $0.9476$\:m and the height is $1.6901$\:m, leading to the grid cell area of $1.6095$\:$m^2$.

\begin{figure*}
        \centering
        \includegraphics[width=7in]{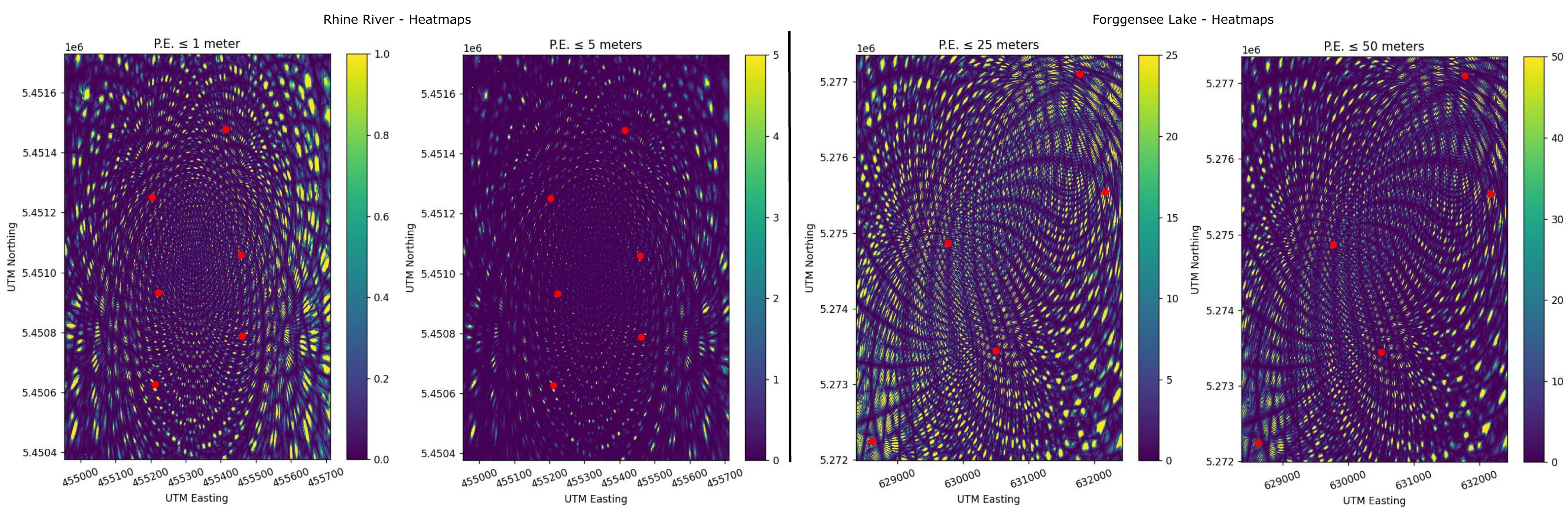}
        \vspace{-12pt}
        \caption{UTM heatmap depictions of the Rhine and Forggensee campaign. MRD locations are marked with red dots. The pattern of granularity is illustrated with respect to positioning error below $1$\:m and $5$\:m for Rhine, $25$\:m. and $50$\:m for Forggensee respectively.}
        \label{fig:heatmaps}
        \vspace{-12pt}
\end{figure*}

\begin{figure*}
    \centering
    \includegraphics[width=7.7in]{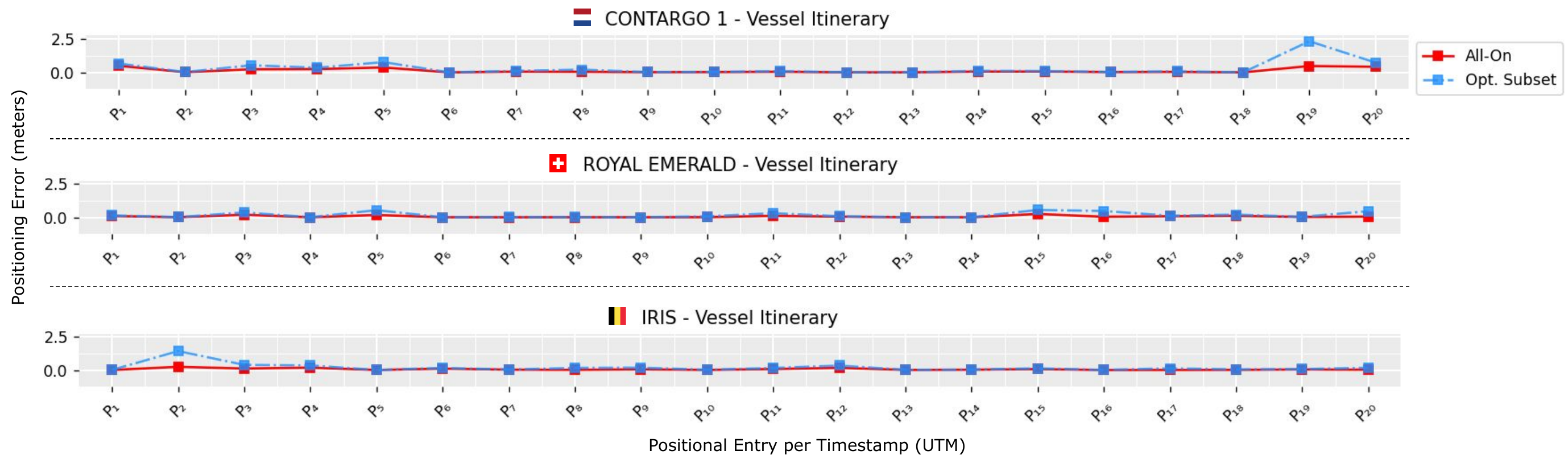}
    \vspace{-9pt}
    \caption{Positioning error measurements for the vessels CONTARGO 1 (NL), ROYAL EMERALD (CH) and IRIS (BE) across distinct sensor combinations.}
    \label{fig:rhine_line_graph}
    \vspace{-12pt}
\end{figure*}

Varying heatmap representations of the region are provided in Fig.~\ref{fig:heatmaps}. The pattern of the given region yield enhanced accuracy not only due to the close proximity of the geography but also due to localization geometry that was utilized amongst MRD receivers. The color legend was modified in order to depict clusters that yield an error of $5$\:m and $1$\:m. Circular clusters that are formed around MRD positions substantiate the derived cosine terms from (\ref{eq:delta_r}). The dispersion of heat values illustrate that for a vessel positioning instance the error value escalates up to the corresponding interval. In order to optimize the expenditure, the system should further detect vessel positions where enhanced accuracy can be obtained without using the entire sensor infrastructure. Three vessels were utilized during the measurement campaign with distinct time series presentation of each nautical course. Fig.~\ref{fig:rhine_line_graph} shows the different uniform sensor combinations regarding three distinct nautical vessels for each timestep (measured on October 3, 2023). With the focus on optimizing the sensor utility, the vessel movement is labeled from $P_1$ to $P_{20}$.

The analysis shows that while there are entries with exorbitant deviation from the mean, the positioning error is below $0.7$\:m. For such a scenario, the optimization is more imminent since operationality of all sensors throughout makes negligible difference to the desired threshold. Most significant deviations from the baseline measurements occurred in positional entries $P_5$, $P_{15}$, $P_{16}$, and $P_{20}$. The values of all combinations are not depicted in the figure. For six sensors, there exist $C(6, 3)$ distinct sensor combinations with 3 sensors. This amounts to $20$ different sensor functionality scenarios. At each time step, the sensor combination that yields the minimum positioning error is taken. Subsequently, the sensor trio that yields the second best accuracy is then detected. 
The best scenario is the one in which all MRD receivers are functioning. 
In certain cases, the angle of a given position is highly optimal that receivers located at the perimeter of the sensor region incorporate negligibly. The values corresponding to $P_5$, $P_{15}$, $P_{16}$, and $P_{20}$ present the highest changes in fractional terms. Compared to the baseline measurements, the highest deviation related to the voyage of Royal Emerald was measured to be $0.4329$ for $P_{16}$. Subsequently, two other nautical ship trajectories were measured during the campaign.

Compared to Royal Emerald, both Iris and Contargo 1 exhibit an increase in the overall positioning error measurements. In particular, average values are under $1$\:m of positioning accuracy, similar to Royal Emerald. However, outliers of Iris are above $1.4$\:m, and outliers of Contargo 1 are almost at $2$\:m. The consistency of baseline measurements are kept. For suboptimal sensor combination solutions, the deviation has increased in comparison with the initial measurements of Royal Emerald. Contrary to Royal Emerald, both Iris and Contargo 1 display much higher disparity between minimum attainable and baseline values. For Iris, this value is observed in entry $P_2$, where the depicted combination has a disparity of $1.1869$\:m between optimal and baseline measurements. For Contargo 1, the maximum disparity is observed in entry $P_{19}$, yielding a value of $1.8722$\:m. The current campaign consists of close proximity sensor emplacements, thus not impeding the overall performance. The modification of emplacement geometry can be evaluated by adding more entries with different emplacement locations. Conclusively, the system is more than capable of performing vessel positioning and optimization schemes along bodies of water with low shoreline proximity.

\subsection{Forggensee Measurement Campaign}
The second testing scenario was conducted in the lake reservoir of Forggensee, north of Füssen, Germany. In contrast to the initial testings carried out at the Rhine river, this scenario presents a lake reservoir where vessels travel in a round-trip fashion. The region covered by the sensors span approximately $2.4$\:km in width and $5.5$\:km in height. Contrary to the initial test region, this serves as an expanded measurement and localization precinct. The assessment of positioning accuracy is evaluated by means of utilizing two passenger ferries that operate between June and October. \textit{MS Allgäu} conducts a small round trip to the south of the lake, between the towns of Füssen and Osterreinen. \textit{MS Füssen} completes a longer route, starting from Füssen all the way up to Roßhaupten.

\begin{figure*}
    \centering
    \includegraphics[width=7.7in]{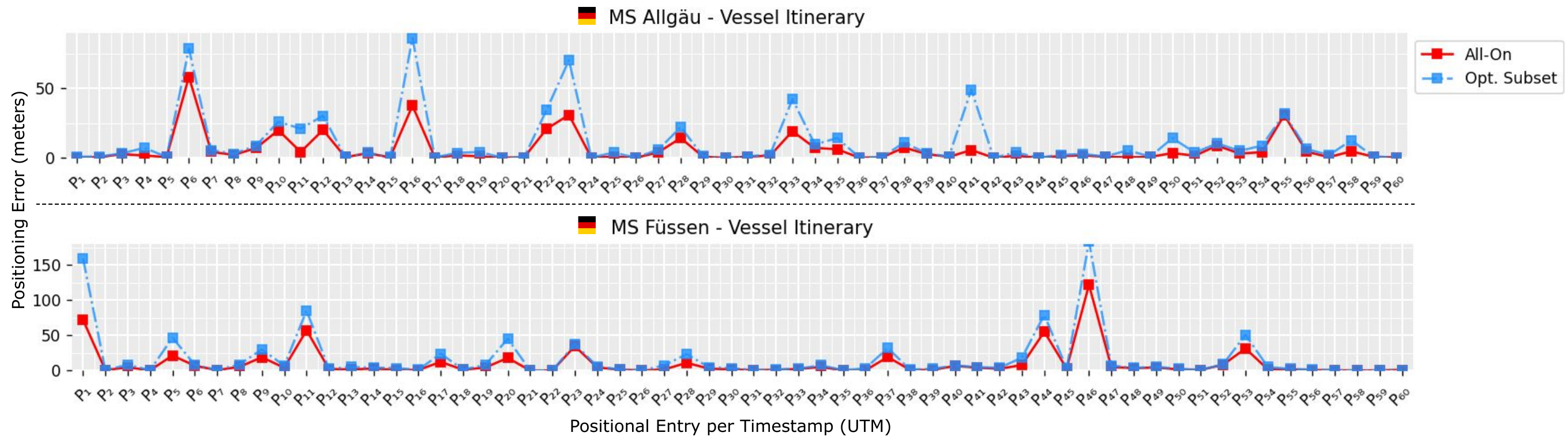}
    \vspace{-9pt}
    \caption{Positioning error estimations of~\textit{MS Allgäu} and~\textit{MS Füssen} with 60 entries from the measurement campaign.}
    \label{fig:rhine_iris_line_graph}
    \vspace{-12pt}
\end{figure*}

Referring to MS Füssen, $35$ distinct measurement instances satisfy the positioning error threshold requirement of $5$\:m. Negligible differences to the baseline measurements are further depicted in numerous entries. Specifically, for entries $P_{21}$, $P_{22}$, and $P_{58}$, the difference between baselines and chosen combinations are on the order of $10^{-4}$. Vessel positioning can be quite good while the incorporation of the farthermost sensor being virtually absent. Apart from scenarios with negligible measurement differences, the actual optimization scenarios can be observed from numerous entries, such as $P_{13}$, $P_{14}$, and $P_{15}$. In the context of MS Allgäu, distinct measurement patterns are similar to MS Füssen but include less outliers pertaining to the baseline measurements. The highest baseline error in accuracy was measured to be $58.3698$\:m for the entry $P_6$. For MS Füssen, it is $121.5958$\:m for the entry $P_{46}$. Results suggest that regions with greater magnitude require more sensors that are active. In contrast with the Rhine river, the region of Forggensee lake has greater distances, which necessitate meticulous configurations of sensor localization geometries with more MRD emplacements.

\section{Conclusion}
\label{sec:conclusion}
In this paper, we developed and evaluated the maritime vessel positioning system and optimized the expenditure related to the operation of this system for distinct tidal regions. The system is aimed as an alternative to GNSS-based positioning systems that are highly vulnerable to jamming and spoofing in the context of nautical navigation and traffic. We looked at the overall performance of the provided positioning system along riverbanks and lakes. Practical scenarios illustrated the proper measurement approach for the maritime vessel radar signals and further utilization within positioning. Optimization schemes pertaining to overall system expenditure were carried out. The contribution of each sensor was evaluated for various vessel-target routes and along the areas with applicable adjacency, and functionality was adjusted accordingly. For other scenarios consisting of larger scales, improvements are required with reference to the number of MRD devices and localization geometries, which is part of our future work. 


\bibliographystyle{IEEEtran}
\bibliography{paper}

\end{document}